\documentclass[10pt,twocolumn]{article}
\usepackage[a4paper,margin=20mm]{geometry}
\usepackage{graphicx}
\usepackage{authblk}
\usepackage{siunitx}
\usepackage{subscript}
\usepackage[hidelinks]{hyperref}
\DeclareSIUnit{\mrad}{mrad}
\newcommand{\bmsection}[1]{\section*{#1}}
\title{Resonant dispersive wave emission: a route toward few-cycle, tunable strong-field drivers in the visible}

\author[1]{Martin Gebhardt}
\author[1]{Joleik Nordmann}
\author[1]{Nikoleta Kotsina}
\author[1]{Michael Heynck}
\author[1]{Teodora Grigorova}
\author[1]{Christian Brahms}
\author[1,2]{John C. Travers}

\affil[1]{School of Engineering and Physical Sciences, Heriot-Watt University, Edinburgh, EH14 4AS, United Kingdom}
\affil[2]{j.travers@hw.ac.uk}

\begin{document}

\twocolumn[
\maketitle
\begin{abstract}
Intense ultrafast lasers are the backbone of strong-field physics and enable table-top scale applications of coherent high-photon energy radiation through high-harmonic generation (HHG). Using short-wavelength drivers in HHG promises highly efficient conversion into the extreme ultraviolet, but short pulse durations are essential to this approach. We generate multi-GW peak power, sub-10 fs pulses in the visible with broad spectral tunability (\qtyrange{220}{752}{\nm}) by harnessing soliton dynamics in large-core, gas-filled capillary fibers for high-energy resonant dispersive wave emission. This establishes the potential of resonant dispersive waves as a new class of driving lasers for HHG and strong-field physics in addition to widespread applications in ultrafast spectroscopy.
\end{abstract}
\vspace{1em}
]

Ultrafast lasers in the ultraviolet (UV) and visible spectral regions are established tools for applications in materials processing \cite{Lee2021} and pump-probe spectroscopy \cite{Kotsina2022}. With ongoing laser performance scaling, there is rapidly growing interest in using intense UV--visible pulses for strong-field experiments, such as high-order harmonic generation (HHG) from noble gas targets \cite{Falco-Filho2010, Wang2015, Marceau2017, Comby2019, Klas2021}.
HHG grants access to highly application-relevant coherent extreme ultraviolet (EUV) light on a table-top scale \cite{Calegari2014, Gardner2017}. However, it suffers from low conversion efficiencies (e.g., $10^{-7}$ from well-established infrared (IR) lasers to the \qty{70}{\eV} photon energy range \cite{Takahashi2004}). The transition to short-wavelength drivers has shown great promise to overcome HHG coversion efficiency bottlenecks \cite{Falco-Filho2010, Wang2015, Marceau2017, Comby2019, Klas2021}, because shorter fundamental wavelengths can be phase-matched to the high-harmonic emission at higher target ionization levels \cite{Durfee1999}. This permits higher driving intensities while maintaining phase matching, which can increase the single-atom emission. Furthermore, faster fundamental field oscillations reduce the quantum diffusion of the electrons accelerated in the process \cite{Schiessl2007}, which leads to higher recombination probability. Given the well-known return kinetic energy scaling with intensity and wavelength ($\sim I\lambda^2$), the main drawback of UV--visible HHG drivers is the limited phase matching cutoff photon energy. However, by employing gases with high ionization potential and using very short driving pulses, the phase-matched intensity can be significantly enhanced, enabling EUV generation beyond \qty{90}{\eV} photon energy based on visible lasers \cite{Cirmi2012, Karst2023}.
Because shorter driving pulses can enhance HHG efficiency \cite{Westerberg2025, Klas2026}, fully exploiting the benefits of visible HHG drivers critically depends on visible ultrafast light sources with very short pulses, i.e., full-width at half-maximum (FWHM) widths below \qty{30}{\fs} \cite{Karst2023}, ideally in the few-cycle regime.

The generation of short, energetic femtosecond pulses in the visible has been demonstrated based on parametric amplification \cite{Wilhelm1997, Odhner2015}, spectral filtering of IR-driven supercontinua \cite{Wirth2011}, IR-driven second harmonic generation (SHG) \cite{Zhou2005, Abdolghader2025}, and nonlinear pulse compression \cite{Liu2010, Descamps2021, Karst2023, Kaufman2025, Liu2026}. Further performance scaling faces a range of challenges associated with the interaction between high-photon energy light and solid-state media. Some of the challenges are limited spectral tunability when frequency-doubling narrowband, fixed-wavelength drivers \cite{Karst2023}, chromatic dispersion and limited crystal thickness \cite{Abdolghader2025}, as well as enhanced nonlinear absorption within the high-index layers of chirped pulse compression mirrors \cite{Razskazovskaya2015, Gui2021}. As an alternative, large-core capillary waveguides filled with noble gas offer excellent transmission and well-controlled dispersion across the visible. Nonlinear intermodal mixing in such waveguides has enabled sub-\qty{5}{\fs} visible pulses, using a 4f grating arrangement to compensate dispersion accumulated after the fiber \cite{Piccoli2021}. Soliton dynamics in these waveguides can generate ultrashort, and crucially, compressed pulses directly \cite{Brahms2020}, without the need for subsequent dispersion compensation. In particular, resonant dispersive wave (RDW) emission via soliton dynamics enables broad UV--visible tunability \cite{Brahms2020} in fundamental mode operation \cite{Travers2019}. Visible RDW generation has been demonstrated at \qty{20} {\uJ}-level energy \cite{Brahms2020}, but further energy scaling and dispersion-minimized separation of the short visible pulses from their driving IR soliton background remain important challenges. Hence, on-target visible RDW peak powers are so far around a few hundred MW \cite{Zhang2022}, which limits their use as strong-field drivers and leads to impractically high phase-matching gas densities in HHG.
\begin{figure*}
\centering
\includegraphics[width=180mm]{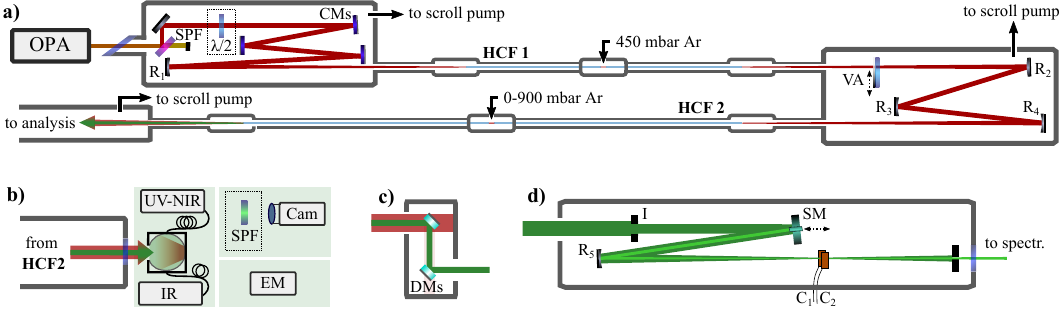}
\caption{Experimental setups of the a) visible RDW generation, b) Spectral, spatial and energy characterization, c) RDW separation and d) the pulse measurement. Sperical mirror curvatures: $R_1=$\qty{-3.5} {\m}, $R_2=$\qty{3.0} {\m}, $R_3=$\qty{-2.0} {\m}, $R_4=$\qty{-4.0} {\m}, $R_5=$\qty{-2.0} {\m}. }
\label{fig1}
\end{figure*}

In this letter, we demonstrate the highest visible RDW energy generated to date (\qty{77}{\uJ}), the highest visible RDW conversion efficiency (15.0\% after correction for input coupling and collimation losses), and high-efficiency isolation of a sub-\qty{10} {\fs} RDW resulting in a peak power $>$\qty{4} {\GW} at \qty{519} {\nm}, a record for visible RDW pulses. This performance opens the door for RDWs to be used as strong-field drivers for the first time. Our results are based on the following key ideas: In previous demonstrations, the conversion efficiency of RDW emission was optimized around \qty{2.5}-\qty{4} fold frequency upconversion \cite{Brahms2020, Travers2019, Yu2025}. Following earlier work \cite{Brahms2020}, we drive soliton dynamics in the \qty{2} {\um} wavelength regime to place this high-efficiency window in the visible. At the same time, the long initial carrier wavelength leads to significant anomalous waveguide dispersion and allows observing soliton dynamics over reasonably short length scales, even at large core dimensions \cite{Brahms2020}. This enables us to scale the RDW energy using large-core stretched capillary setups \cite{Travers2019}. Finally, we use a combination of spatial filtering (the free-space divergence of the fundamental fiber mode scales as $\lambda$) and dispersion-minimized dielectric mirrors to isolate a near transform-limited RDW in the green spectral range at high energy.

The setup is shown in Fig. \ref{fig1}. Our experiments are driven by a commercially available optical parametric amplifier (OPA) and all free-space propagation sections (including the pulse characterization setups) are housed in vacuum chambers which we evacuate with a set of scroll pumps. The OPA output is sent into a vacuum chamber where we isolate the idler using a short-pass filter (SPF). The idler pulses have a passively stabilized carrier-envelope phase (CEP), are p-polarized, and are centered at \qty{1830} {\nm} with a bandwidth supporting \qty{31} {\fs} Fourier transform-limited (FTL) FWHM pulse duration. The maximum pulse energy is \qty{3.7} {\mJ} at 100 Hz repetition rate. To enable efficient RDW emission, we pre-compress the pulses through spectral broadening in a first hollow capillary fiber (HCF 1, \qty{700} {\um} core diameter, \qty{2}{\m} length).

We find that coupling positively chirped \qty{60} {\fs} FWHM pulses to HCF 1 (using chirped mirrors, CMs) allows us to avoid pressure-dependent transmission loss, as is otherwise observed under the influence of ionization. HCF 1 is filled with argon at \SI{450}{\milli\bar} pressure from a central access point to the core as shown in Fig. \ref{fig1}a)~\cite{Kotsina2026}. With the optical input and output held at sub-mbar level, the pressure profile in the waveguide follows a well-defined gradient \cite{Suda2005} toward both ends. This method avoids propagation of the high-energy pulses through any additional windows. In this configuration we are able to generate \qty{2.0} {\mJ}, \qty{10} {\fs} pulses after external compression (see Supplement). To drive high-energy RDW generation, we pass the spectrally broadened HCF 1 output through a variable transmissive attenuator (VA, Thorlabs NDL-25C-2, see Fig. \ref{fig1} a). This component allows for compact energy control and dispersion compensation. Its GDD of \qty{-128} {\fs^2} at \qty{1.8} {\um} wavelength is sufficient to achieve temporal compression of the pulses toward the end of HCF 2, which itself provides non-negligible anomalous waveguide dispersion of \qty{-12.3} {\fs^2 m^{-1}} at \qty{1.8} {\um} wavelength. HCF 2 is a \qty{3.7} {\m} long, \qty{700} {\um} core diameter stretched capillary with a central pressure control point. We achieve fundamental mode excitation with a set of spherical mirrors (Fig. \ref{fig1}). The experimental transmission through this fiber is as high as 84\%, while the theoretical limit is 88\%.

We first demonstrate the overall spectral tuning and energy scalability of our approach by optimizing the input pulse energy and pressure set point in HCF 2 for the brightest RDW emission. The fiber output is characterized directly after a MgF\textsubscript{2} window as shown in Fig. \ref{fig1} b. The reported output energies are not corrected for the window transmission. We use a single integrating sphere fiber-coupled to separate UV--NIR and IR spectrometers to measure the spectral energy density; an energy meter (EM) for the overall output; and an optional \qty{750} {\nm} SPF followed by a camera (Cam) to characterize the beam. The complete spectral measurement system is calibrated using an absolute reference at the sphere entrance. At the maximum HCF 2 output pulse energy of \qty{1.3} {\mJ} and a pressure of \SI{85}{\milli\bar} we observe a broad supercontinuum spanning the UV to IR and a weak RDW peak around \qty{220} {\nm}. As we increase the argon pressure and appropriately reduce the input pulse energy, we are able to tune the phase-matched RDW emission across the ultraviolet, visible, and near-infrared regions. The result of this measurement is presented in Fig. \ref{fig2}, with the full details listed in the supplementary information. We are able to generate substantial pulse energies in the UV (over \qty{60} {\uJ}), and we measure >\qty{75} {\uJ} of visible pulse energy integrated over a wavelength window extending $\pm20\%$ around the RDW peak wavelength. The conversion efficiency is calculated with respect to the HCF 2 input energy (measured by reflecting the beam after $R_3$ out of the vacuum chamber). This efficiency is not corrected for coupling efficiency or propagation loss in the waveguide. The efficiency rises from the deep UV into the visible, exceeding 10\% at \qty{500} {\nm}.

\begin{figure}[ht!]
\centering\includegraphics[width=82mm]{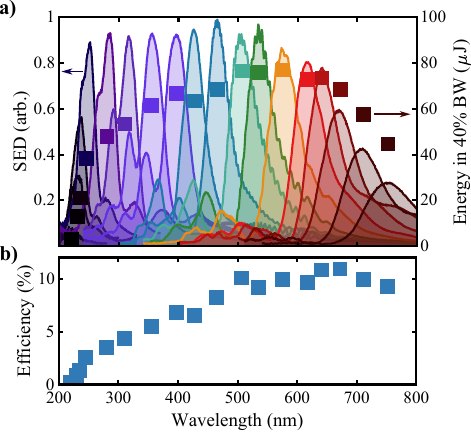}
\caption{RDW tuning spectra, together with integrated pulse energy (a) and corresponding conversion efficiencies (b) for HCF 2 argon pressures ranging from \SI{85}{\milli\bar} to \SI{887}{\milli\bar}.}
\label{fig2}
\end{figure}

\begin{figure}[ht!]
\centering\includegraphics[width=81mm]{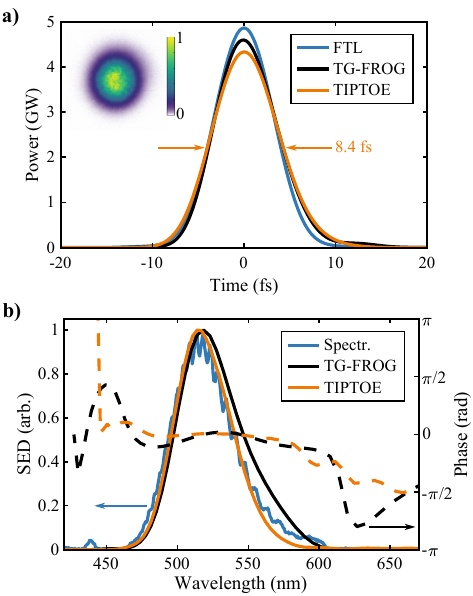}
\caption{Temporal (a) and spectral (b) characterization of \qty{39} {\uJ} energy RDW pulses. Inset: visible beam profile after HCF 2.}
\label{fig3}
\end{figure}

Next, we demonstrate efficient separation of a visible RDW from its IR background in a configuration with modified polarization and IR input chirp. We collimate the broadband HCF 2 output using an off-axis reflective telescope. The collimated $1/\mathrm{e}^2$ beam diameter is about \qty{6} {\mm} in the \qty{500} {\nm} wavelength range, while the \qty{1.8} {\um} fundamental beam is more than three times larger. Thus, it is significantly clipped by the clear aperture of a set of $45^{\circ}$ angle of incidence 1-inch diameter dielectric mirrors (DMs, see Fig. \ref{fig1} c), as well as by the subsequent \qty{12} {\mm} clear aperture of our characterization setup. The DMs are optimized for operation in s-polarization and provide high reflectivity (HR) as well as low GDD within a bandwidth spanning \qty{450}-\qty{610} {\nm}. Outside of this HR window, the average theoretical reflectivity is about 15\% per bounce with strong GDD oscillations. For the RDW filtering experiments we modify the setup by inserting a half-wave plate into the beam path before HCF 1 which changes the polarization throughout the beamline to s-pol. Additionally, its additive negative GDD in the IR leads to short, slightly negatively chirped pulses at the input of HCF 2. We find that this adjustment allows observing RDW emission at reduced energies, and that the RDW generation bandwidth is slightly narrower. Hence, as compared to the RDW spectral tuning (Fig. \ref{fig1}), we achieve optimized conversion efficiency into the HR mirror window. When the HCF 2 central pressure is \SI{450}{\milli\bar} and the input energy is set to \qty{344} {\uJ}, the RDW central wavelength is \qty{519} {\nm}, which aligns well with the HR mirror band. We then measure the broadband spectra with our calibrated spectrometers before and after the DM setup (see Supplement). The filtered RDW energy is \qty{39} {\uJ} with an additional \qty{6} {\uJ} remaining in the UV and IR. These numbers are in reasonable agreement with the EM measurement of \qty{43} {\uJ} in total after both DMs. Therefore, the conversion efficiency from the IR pulse energy incident on HCF 2 to the visible RDW energy after collimation and separation is 11.3\%. When corrected for the collimation optics as well as the HCF 2 input coupling, the efficiency reaches 15.0\%. At this point, the HCF 2 output beam profile measured after a \qty{750} {\nm} SPF shows a symmetric Gaussian-like beam profile as is expected in fundamental mode operation (see inset Fig. \ref{fig3}).

After separation, we send the green pulses to a home-built in-vacuum pulse characterization setup (Fig. \ref{fig1} d), which is based on TIPTOE \cite{Park2018,Larsen2025}. Following an iris (I), the pulses are split into two time-delayed replicas by an annular segmented mirror (SM), and focused between two electrodes C\textsubscript{1,2}. The measurement signal is created within the \qty{140} {\um} focus by ionization of the background argon gas. To cross-check our results, we further measure the separated RDW pulses with an in-vacuum transient-grating frequency-resolved gating (TG-FROG) setup. The pulses propagate from HCF 2 to both temporal diagnostics entirely in vacuum, without passing through the MgF\textsubscript{2} window used for the tuning scan. Fig. \ref{fig3} presents the measurement results in the time (a) and spectral (b) domains. The TIPTOE measured RDW has a FWHM duration of \qty{8.4} {\fs} and its retrieved spectrum agrees well with both the TG-FROG result and an independent spectrometer measurement. Normalizing the TIPTOE-retrieved temporal intensity profile to the measured RDW energy of \qty{39} {\uJ}, we estimate an on-target peak power of \qty{4.3} {\GW}, which is close to the transform limit (\qty{7.7} {\fs} FWHM duration, \qty{4.9} {\GW} peak power).

\begin{figure}[ht!]
\centering\includegraphics[width=82mm]{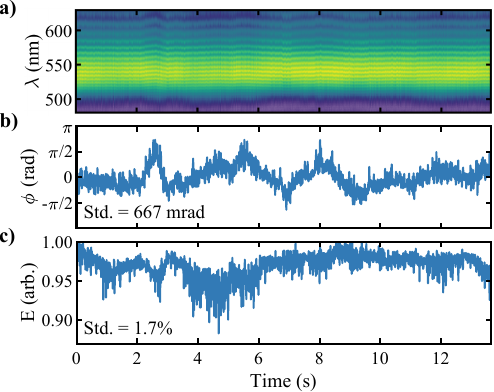}
\caption{Measured RDW single-shot spectra (a) and retrieved fringe phase fluctuations (b), together with RDW energy stability measured using a single-shot energy meter (1.7\% relative standard deviation) over 1350 laser shots.}
\label{fig4}
\end{figure}

As a final step, we analyze shot-to-shot RDW spectrum measurements acquired over $>\qty{10}{\s}$ (Fig. \ref{fig4} a). We observe spectral fringes that originate from interference between the RDW and a weak, non-phase-matched third harmonic (TH) of the IR. The TH is generated in HCF 1 and appears predominantly in higher order spatial modes (leading to low fringe-visibility). However, it is passed through the broadband setup allowing us to infer the CEP jitter. Although the CEP in a true $f$--$3f$ interference scheme follows half of the spectral fringe movement \cite{Ishizawa2022}, we cannot decouple phase and amplitude drifts originating from separate dynamics in HCF 1 and 2. Hence, we use the directly retrieved spectral fringe phase $\phi$ as upper bound for the RDW CEP. Its standard deviation is \SI{667} {\mrad}, and the main fluctuations are observed on a $\sim$\qty{1} {\s} timescale (Fig. \ref{fig4} b). We attribute these drifts to slow amplitude and phase fluctuations of the OPA idler, as well as to slow pointing drifts before HCF 1. There is a weak correlation between the  slow RDW energy drift (E, Fig. \ref{fig4} c) and $\phi$, however, it is important to note that rapid shot-to-shot CEP fluctuations are well below \SI{300} {\mrad}.

In conclusion, we demonstrate visible RDW generation with record pulse energies of \qty{77}{\uJ} and conversion efficiencies above 10\%. In a configuration optimized for separation, we isolate \qty{39}{\uJ} pulses at \qty{519}{\nm} with an \qty{8.4}{\fs} duration and an estimated on-target peak power of \qty{4.3}{\GW}, without additional dispersion compensation. These advances make high-energy RDW emission relevant for HHG and other strong-field experiments.

In particular, the short pulse durations achieved here represent a key tool to study the efficiency scaling of visible-driven HHG (currently $>10^{-4}$ into a single harmonic at \qty{26.5}{\eV} in krypton with \qty{20} {\fs} pulses \cite{Klas2021}), with prospects for substantial improvements at shorter pulse widths \cite{Westerberg2025,Klas2026}. Furthermore, intense sub-\qty{10} {\fs} pulses in the visible will allow for full phase matching to >\qty{90} {\eV} photons in helium targets. Our results showcase another appealing feature of RDW emission: its spectral tunability. This allows setting the optimum driving wavelength for a specific HHG photon energy and can be used for spectral tuning of the high harmonic comb lines. Our results suggest that high-energy RDWs can be phase stabilized to a standard deviation in the few \SI{100} {\mrad} range, making it a new candidate to drive attosecond science and field-sensitive strong-field phenomena.

Future work will focus on demonstrating RDW-driven HHG and scaling UV--visible RDW generation~\cite{Kotsina2026}. Broadband, low-dispersion separation optics could retain more of the RDW energy and bandwidth. The spectra in Fig. \ref{fig2}(a) support Fourier transform-limited pulse durations below \qty{6}{\fs} ($<3.6$ cycles at \qty{500}{\nm}) \cite{Brahms2020}. Further geometrical scaling could increase pulse energies, and power-scalable thulium-doped fiber lasers \cite{Wang2023} could enable higher repetition rates. Combined energy and bandwidth scaling offers a route to peak powers in the tens of GW.

\bmsection{Funding} This work was funded by the European Research Council (ERC) under the European Union's Horizon 2020 research and innovation programme: Consolidator Grant agreement XSOL no.~101001534. M.G. acknowledges support by the UKRI funding guarantee scheme for Marie Sklodowska-Curie Actions (EP/Z001250/1). J.C.T. is supported by a Chair in Emerging Technology from the Royal Academy of Engineering and the Institution of Engineering and Technology (IET) through the IET A F Harvey Engineering Research Prize. C.B. is supported by Royal Academy of Engineering Research Fellowship RF/202122/21/133 and the European Research Council through Starting Grant agreement FASTER, 101161675.

\bmsection{Acknowledgment} The authors would like to thank Mohammed Sabbah, Balazs Plosz and Deepjyoti Satpathy for fruitful discussions. 

\bmsection{Disclosures} The authors declare no conflicts of interest.

\bmsection{Data Availability Statement} Data may be obtained from the authors upon request.

\bibliography{export}
\end{document}